\documentclass{JAC2003}

\usepackage{graphicx}
\usepackage{booktabs}
\usepackage{subfigure}

\begin{document}
\title{A STUDY OF BEAM ALIGNMENT BASED ON COUPLING MODES IN THIRD HARMONIC SUPERCONDUCTING CAVITIES AT FLASH\thanks{Work supported in part by the European Commission under the FP7 Research Infrastructures grant agreement No.227579.}\\[-.8\baselineskip]}

\author{P.~Zhang$^{1,2,3,}$\thanks{pei.zhang@desy.de}, N.~Baboi$^2$ and R.M.~Jones$^{1,3}$\\
\mbox{$^1$School of Physics and Astronomy, The University of Manchester, Manchester, U.K.}\\
\mbox{$^2$Deutsches Elektronen-Synchrotron (DESY), Hamburg, Germany}\\
\mbox{$^3$The Cockcroft Institute of Accelerator Science and Technology, Daresbury, U.K.}}

\maketitle

\begin{abstract}
This paper presents a study of beam alignment in accelerating cavities based on beam-excited higher order modes (HOM). Among the transverse HOMs, dipole modes have the dominant contribution, and may have an adverse ef\mbox{}fect on the beam quality. Since they depend linearly on the transverse beam of\mbox{}fset, their ef\mbox{}fect can be reduced by aligning the beam in the cavity. The study has been made on the four third harmonic superconducting cavities installed in the ACC39 module at FLASH. Modes with strong beam coupling were monitored while moving the beam and searching for the minimum excited HOM power. 
\end{abstract}

\section{INTRODUCTION}\label{sec:intro}
An electron beam excites higher order modes (HOMs) when passing through an accelerating cavity. The transverse components are dominated by dipole modes \cite{rwake}. These can def\mbox{}lect the beam and therefore dilute the beam quality and can potentially lead to a beam instability. Since their strength depends linearly on the transverse beam of\mbox{}fset from the cavity axis, the adverse ef\mbox{}fect can be reduced by aligning the beam on the axis \cite{racc1-2}.

\begin{figure}[h]
\centering
\includegraphics[width=0.47\textwidth]{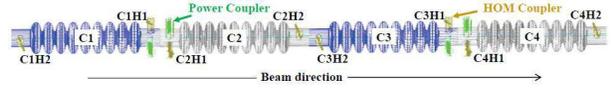}
\caption{Schematic of the four cavities within ACC39. The power couplers (green) are placed downstream for C1 and C3, and upstream for C2 and C4. The HOM couplers (brown) located on the same side of the power couplers are named H1, while the other ones H2.}
\label{cavity-cartoon}
\end{figure}

At FLASH \cite{rflash}, the electron beam is accelerated using TESLA cavities operating at 1.3~GHz. This induces a non-linear energy spread in the bunch compression process, which is corrected by using third harmonic cavities operating at 3.9~GHz \cite{racc39}. There are four 3.9~GHz cavities built in the cryo-module ACC39 (Fig.~\ref{cavity-cartoon}). By design, the 3.9~GHz cavities have two features. First, the size of the 3.9~GHz cavity is three times smaller than that of the 1.3~GHz cavity. This makes the HOMs stronger in the third harmonic cavity \cite{rwake}. Second, the beam pipes connecting the 3.9~GHz cavities are larger than one third of the TESLA cavities. Thus the frequency of most HOMs is above the cut-of{}f frequency of the beam pipes, and therefore are able to propagate through the entire module.

\section{COUPLING MODES IN THE SECOND DIPOLE PASSBAND}\label{sec:simu-meas}
Due to their strong couplings to the beam, cavity modes in the f\mbox{}irst two dipole passbands are the natural choice to use for beam alignment. These modes couple to adjacent cavities through attached beam pipes. This is shown by both simulations and measurements.

\begin{figure}[h]\center
\includegraphics[width=0.48\textwidth]{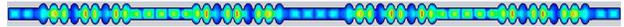}
\caption{Electric f{}ield distribution of the strongest coupling mode in the second dipole passband. The frequency is 5.4192~GHz and the $R/Q$ is 50.92~$\Omega/$cm$^2$.}
\label{simu-full}
\end{figure}

Simulations are performed on a string of four ideal cavities without couplers using CST Microwave Studio \cite{rcst}, with a solver accuracy of 10$^{-6}$ in terms of the eigensystem's relative residual, 1.1~million mesh cells and electric boundary conditions. The resulting f{}ield distribution of the strongest coupling mode in the second dipole passband is shown in Fig.~\ref{simu-full}. The propagation of the electromagnetic energy across the entire cavity string can be clearly seen.

\begin{figure}[h]
\centering
\subfigure[Transmission spectra of C1]{
\includegraphics[width=0.48\textwidth]{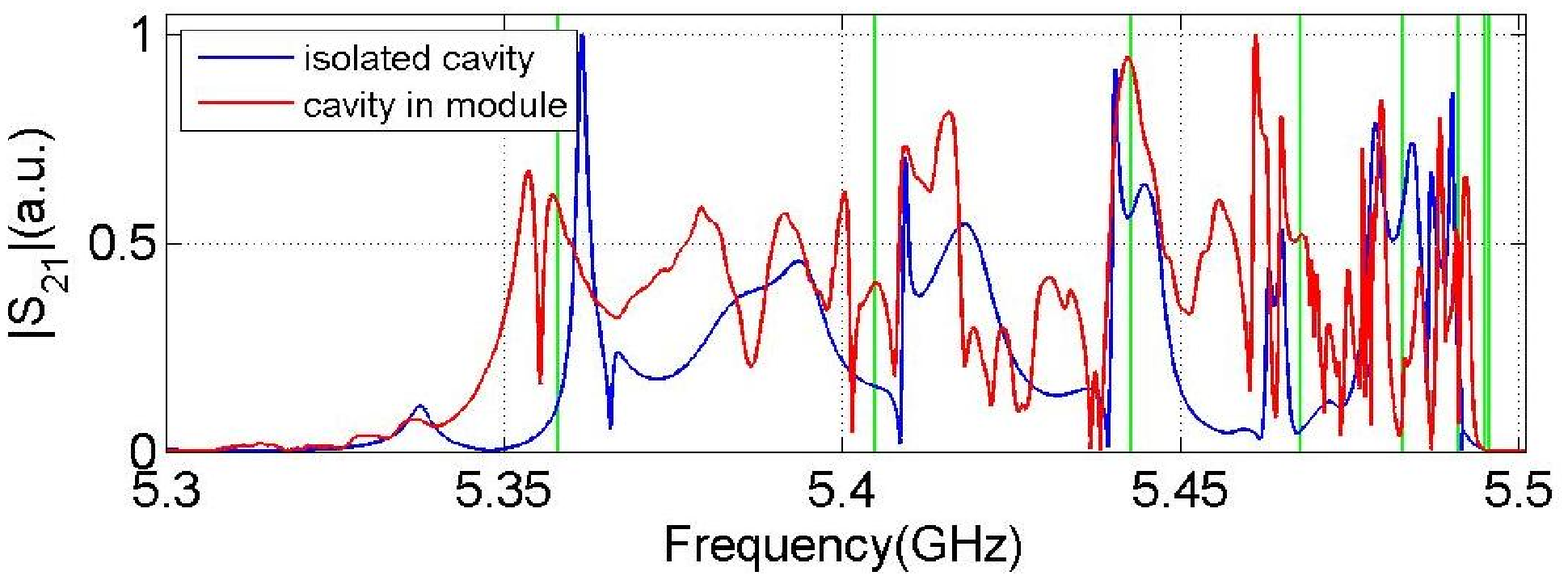}
\label{cmtb-fnal-D2-C1}
}
\subfigure[Transmission spectra from module-based measurement]{
\includegraphics[width=0.48\textwidth]{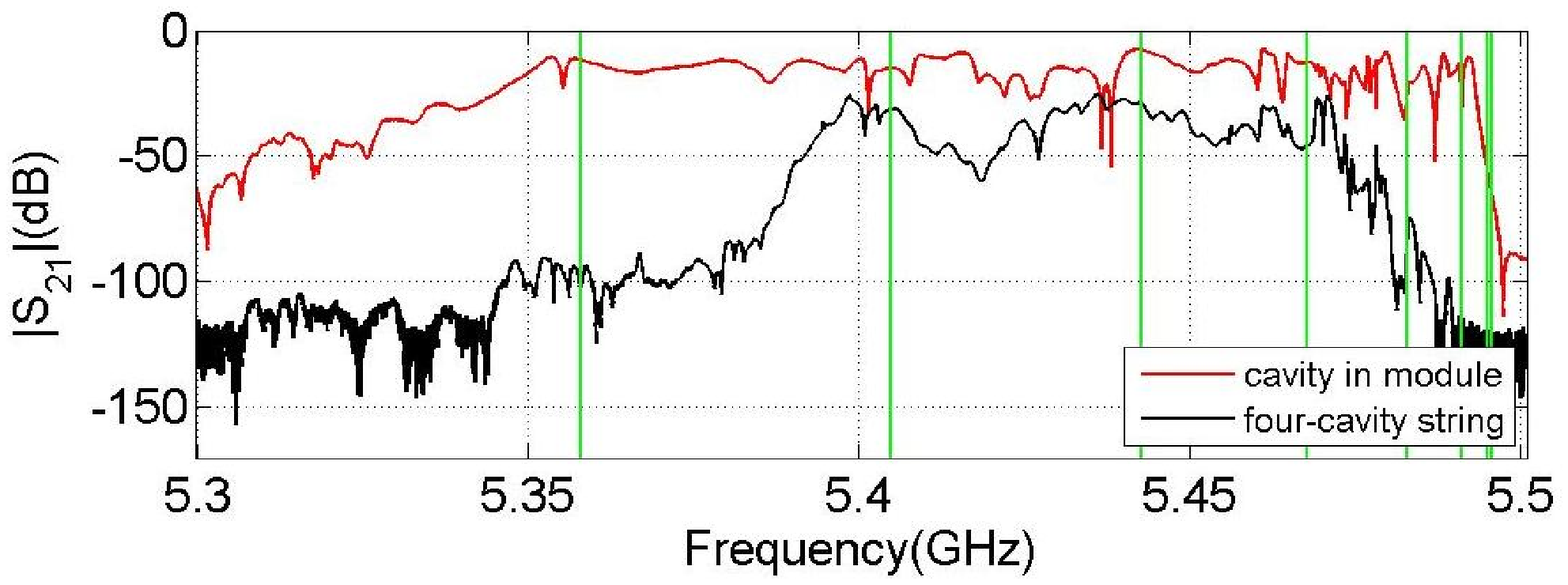}
\label{cutoff-C1}
}
\caption{The second dipole passband from the transmission measurements. The vertical lines in green are from simulations of an ideal cavity.}
\label{fnal-cmtb-cutoff-D2-C1}
\end{figure}

Fig.~\ref{cmtb-fnal-D2-C1} shows the spectrum of C1 measured in isolated cavity case compared with that measured when cavities were assembled in the cryo-module. Because of the coupling ef\mbox{}fects, more modes are present. Fig.~\ref{cutoff-C1} shows the spectrum of C1 measured in the cryo-module together with that measured across the entire four-cavity string (from C1H2 to C4H2). Most modes in the second dipole passband propagate. The dipole character of this passband has been studied in \cite{rdipac}. 

As modes close to 5.45~GHz have strong coupling to the beam, the spectrum from 5.42 to 5.45~GHz is used for the beam alignment study.

\section{Beam Alignment}\label{sec:beam-align}
\subsection{Measurement Scheme}\label{sec:meas-setup}
\begin{figure}[h]
\centering
\includegraphics[width=0.46\textwidth]{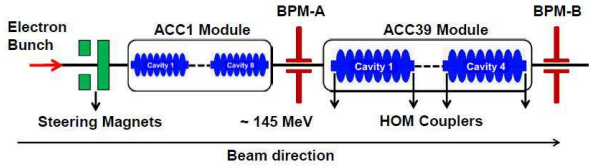}
\caption{Schematic of measurement setup (not to scale).}
\label{hom-setup}
\end{figure}

Aligning the beam corresponds to sending the beam on a trajectory which generates minimum HOM power in the cavities. To this end, we set up the measurement as shown in Fig.~\ref{hom-setup}. Two steering magnets were used to kick the beam horizontally and vertically. The bunch consisting of a charge of approximately 0.5~nC was subsequently accelerated by ACC1 which contains eight 1.3~GHz TESLA superconducting cavities, then went through ACC39 with various transverse of\mbox{}fsets. The position within each cavity is obtained by interpolating the readouts from two beam position monitors (BPM-A and BPM-B), which are situated on each side of ACC39. A straight-line trajectory of the beam was attained by switching of{}f the accelerating f{}ield in ACC39 and the quadrupole magnets nearby. 

\subsection{Minimization of HOMs}\label{sec:min-hom}
\begin{figure}[h]
\centering
\subfigure[Steering magnet]{
\includegraphics[width=0.226\textwidth]{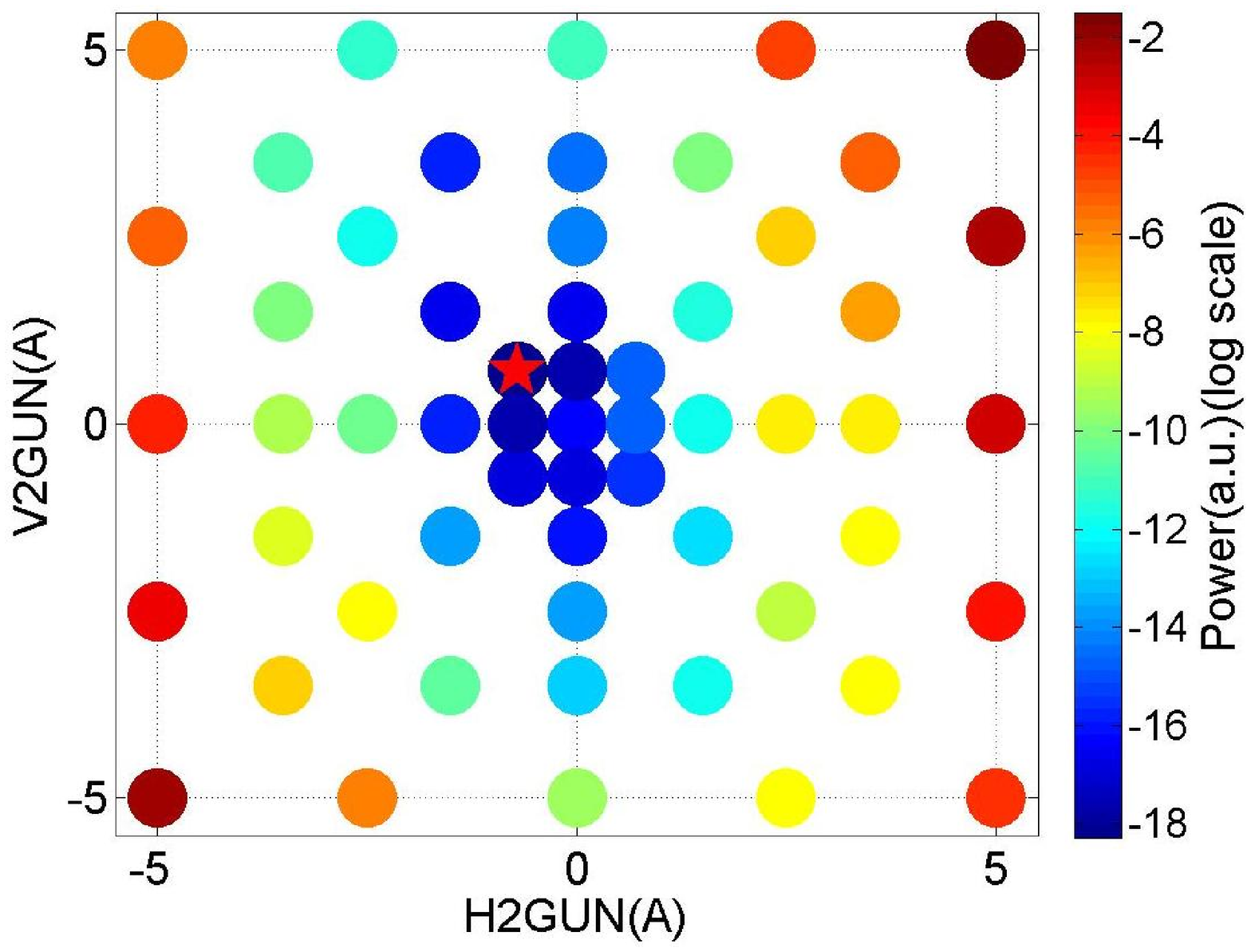}
\label{scan-magnet-old}
}
\subfigure[Interpolated position(C2H2)]{
\includegraphics[width=0.226\textwidth]{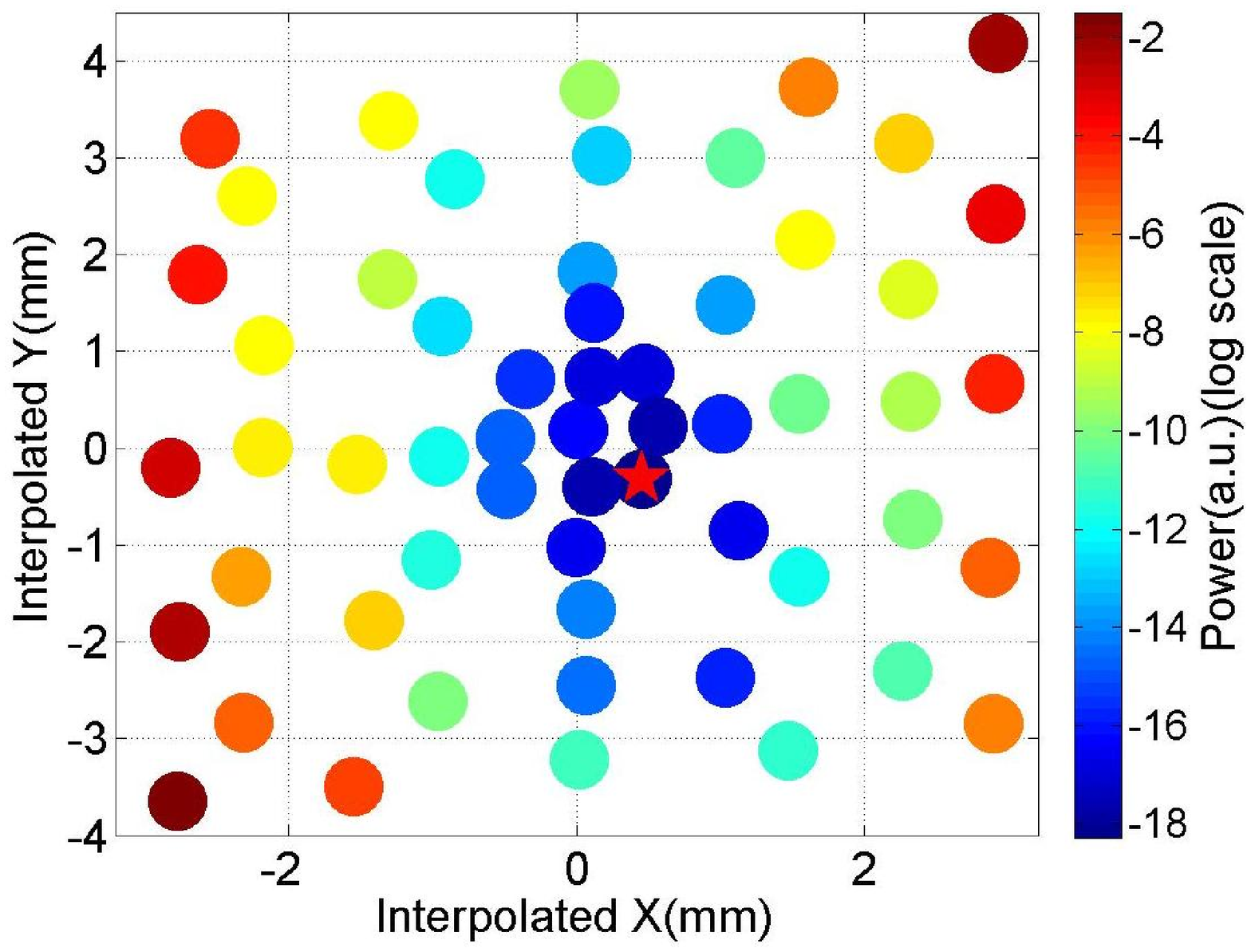}
\label{scan-xy-old}
}
\caption{HOM power measured from C2H2 for each beam position. The position marked with red pentagram has the minimum power.}
\label{scan-min-old}
\end{figure}

From previous measurements \cite{rdipac}, the integrated power of HOM spectrum from 5.42 to 5.45~GHz has a minimum during the grid-like beam movement. Fig.~\ref{scan-magnet-old} shows the integrated power for each beam position in terms of steering magnet current. Position interpolations from the two BPM readouts (BPM-A and BPM-B) are applied to get the transverse beam positions in C2, which is shown in Fig.~\ref{scan-xy-old}. The nonlinearity of the transverse positions during the scan is due to the coupling between $x$ and $y$ plane caused partially by the ACC1 module and partially by BPMs. The jitters of the BPM reading during the beam movement can also contribute. 

\begin{figure}[h]
\centering
\subfigure[Steering magnet]{
\includegraphics[width=0.226\textwidth]{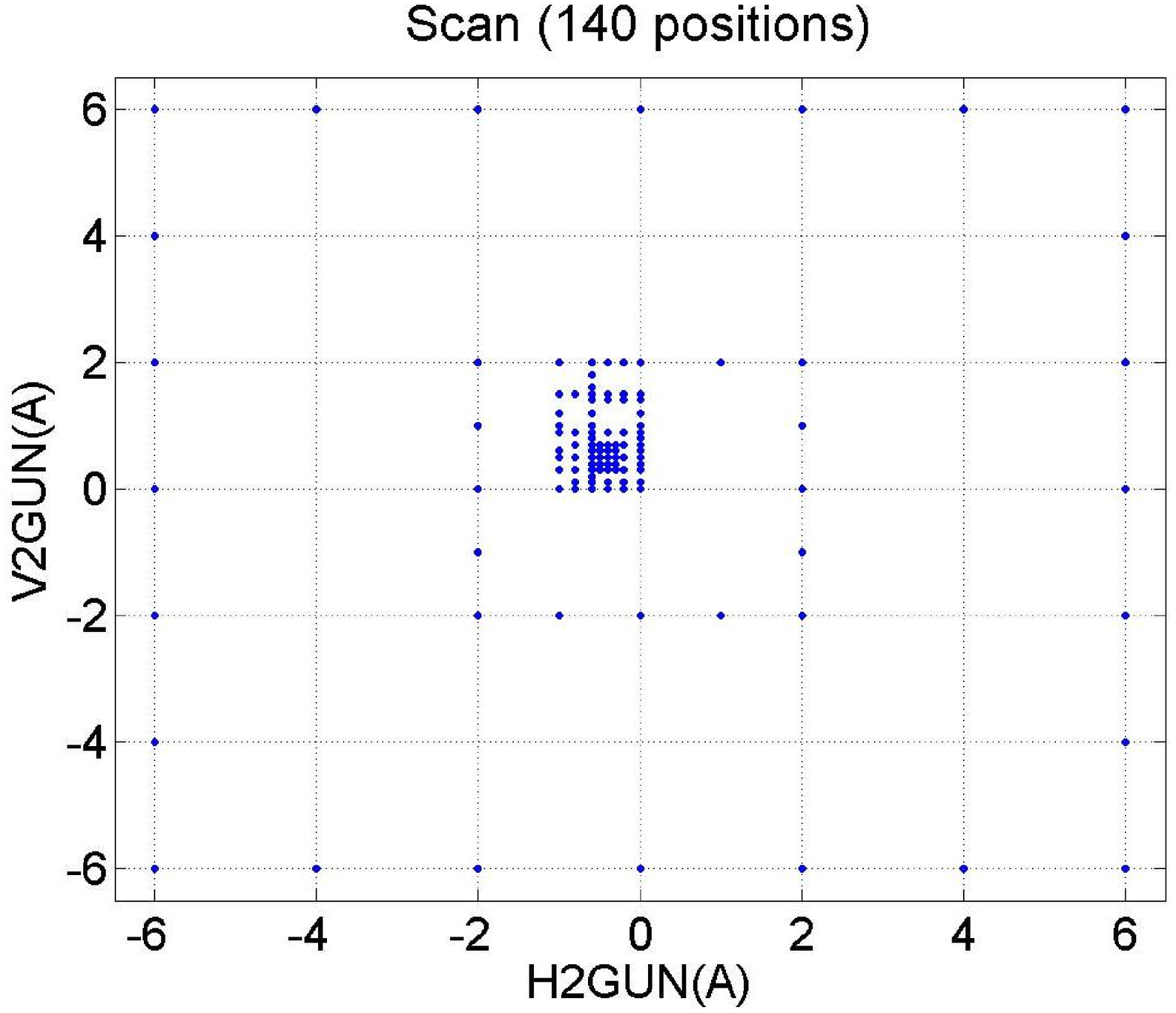}
\label{2D-magnet}
}
\subfigure[BPM readout]{
\includegraphics[width=0.226\textwidth]{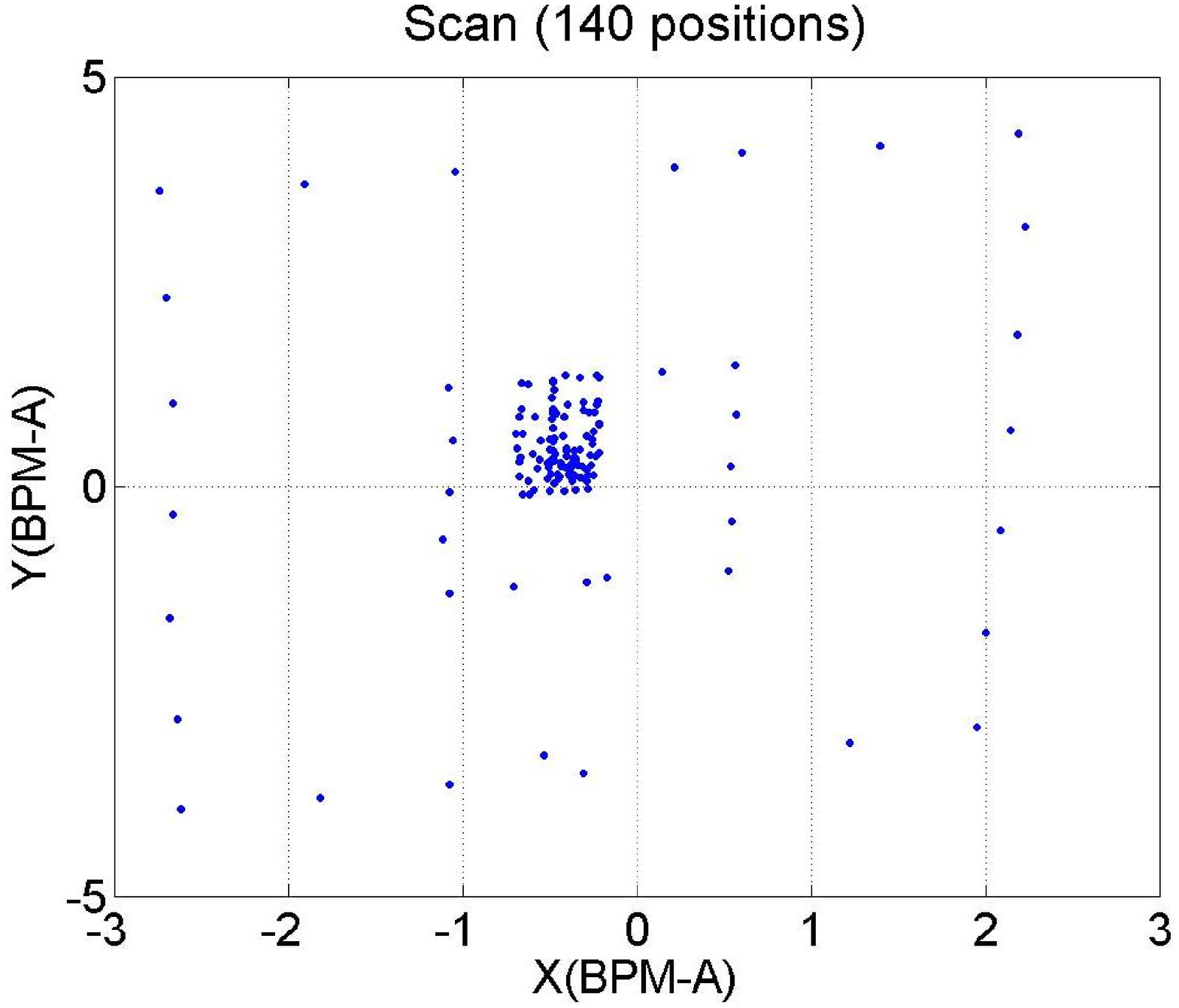}
\label{2D-bpma}
}
\caption{Beam movement.}
\label{2D-magnet-bpm}
\end{figure}

Beam alignment was conducted by minimizing the HOM power measured from coupler C2H2, while monitoring the signals from other couplers as well. Initially, transverse of\mbox{}fsets read from BPM-A as large as 2.7~mm ($x$) and -4.3~mm ($y$) were used and they were subsequently reduced down to -200~$\mu$m ($x$) and -30~$\mu$m ($y$) according to HOM power. Fig.~\ref{2D-magnet} shows the current readings of the two steering magnets during the beam movement, and Fig.~\ref{2D-bpma} shows the corresponding BPM readouts. The nonlinearity presented is caused by the same reasons as described for Fig.~\ref{scan-xy-old}.

\begin{figure}[h]
\centering
\includegraphics[width=0.48\textwidth]{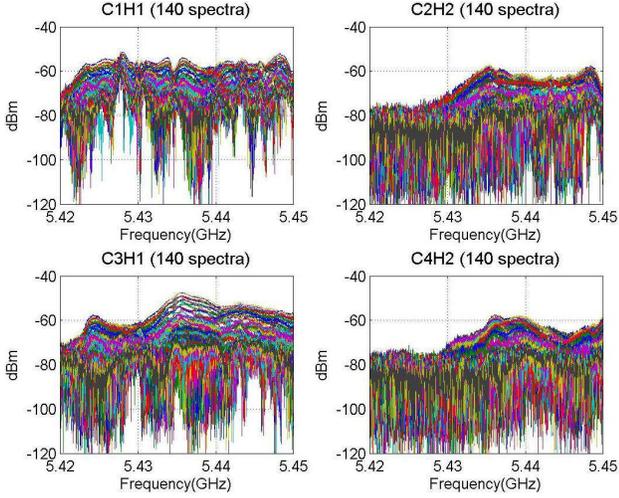}
\caption{Spectra measured from all four downstream HOM couplers for each beam trajectory.}
\label{rsa-overlap-4couplers}
\end{figure}

\begin{figure}[h]
\centering
\subfigure[Steering magnet]{
\includegraphics[width=0.226\textwidth]{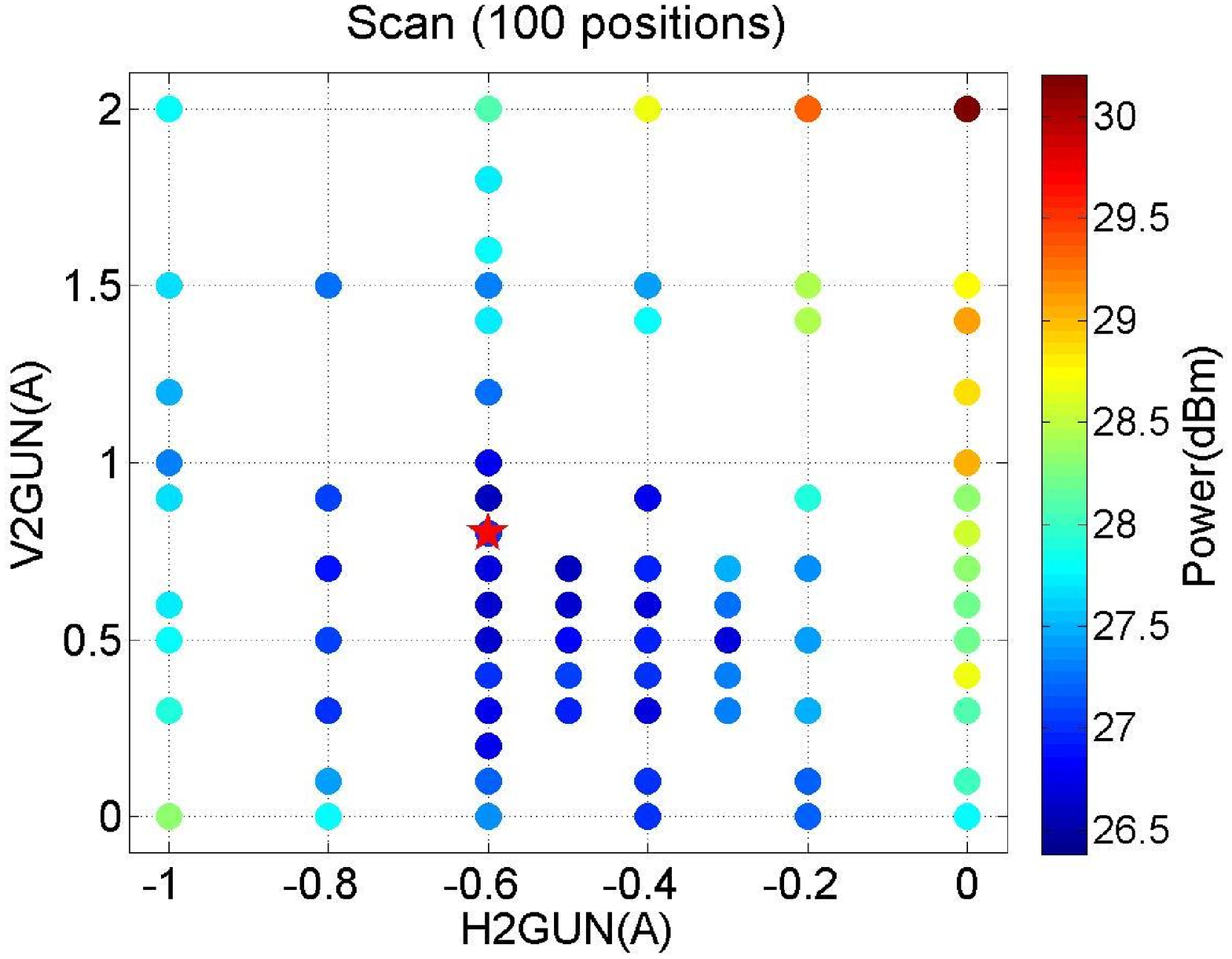}
\label{scan-magnet}
}
\subfigure[Interpolated position(C2H2)]{
\includegraphics[width=0.226\textwidth]{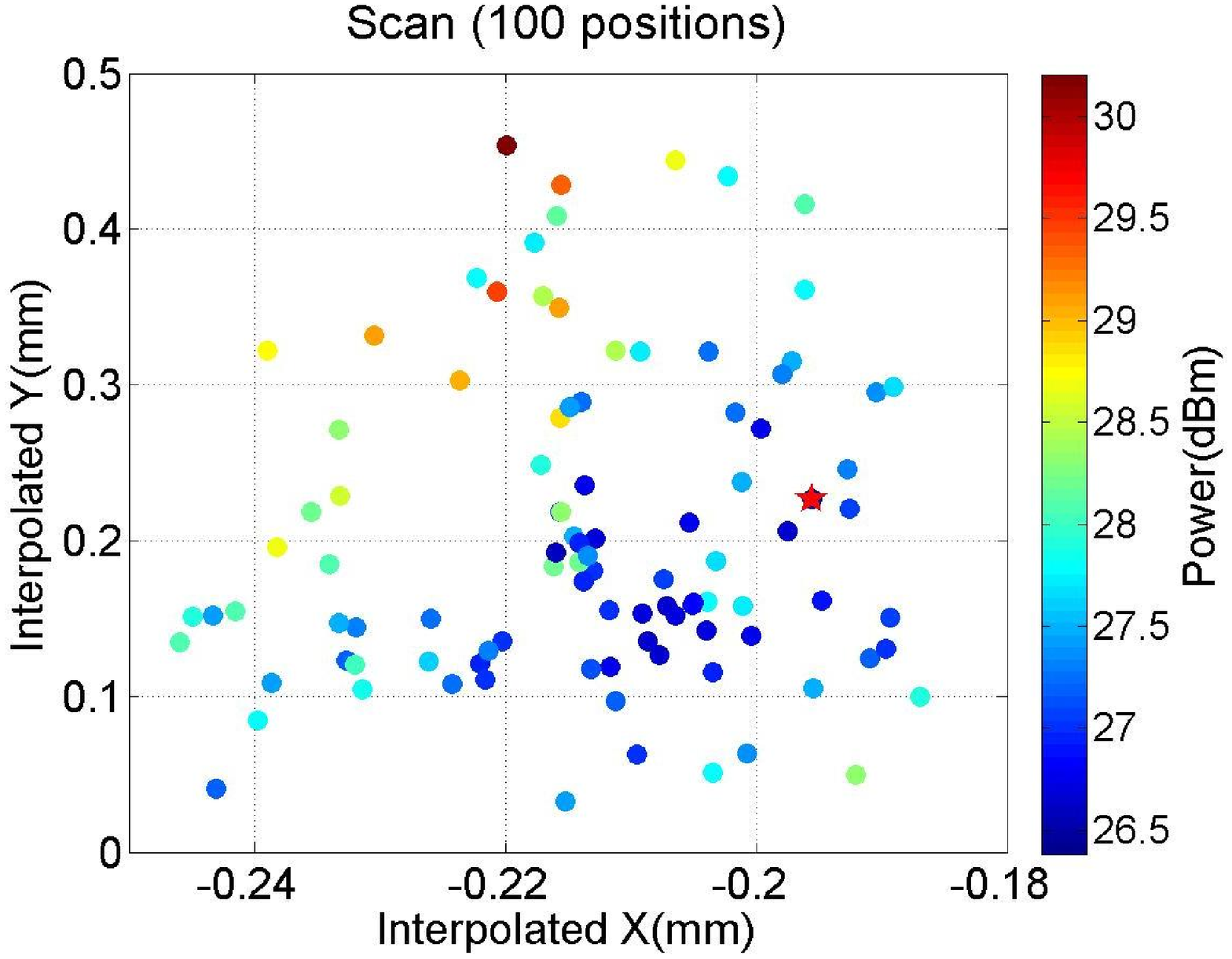}
\label{scan-xy}
}
\caption{HOM power for each beam position. The position marked with red pentagrams corresponds to the minimum HOM power.}
\label{scan-min}
\end{figure}

The HOM signals excited by a single electron bunch for each beam position were measured by a Real-time Spectrum Analyzer (RSA) from all four downstream HOM couplers. The mode amplitude varies with transverse beam of{}fset in each cavity, which is shown in Fig.~\ref{rsa-overlap-4couplers}. Due to the coupling ef{}fects, the spectra are too complex to identify single modes, therefore, the integrated power within this 30~MHz range was used for beam alignment. Zooming into a smaller range of beam movement, the integrated power from C2H2 for each beam position is shown in terms of steering magnet current (Fig.~\ref{scan-magnet}) and interpolated beam position (Fig.~\ref{scan-xy}) respectively. The position which has minimum power from C2H2 is found and marked as red pentagram. The HOM power within a position range of 60~$\mu$m ($x$) and 200~$\mu$m ($y$) in C2 presents very small changes.

\subsection{Beam Trajectory of Minimum HOM Power}\label{sec:opt-traj}
\begin{figure}[h]
\centering
\subfigure[x]{
\includegraphics[width=0.226\textwidth]{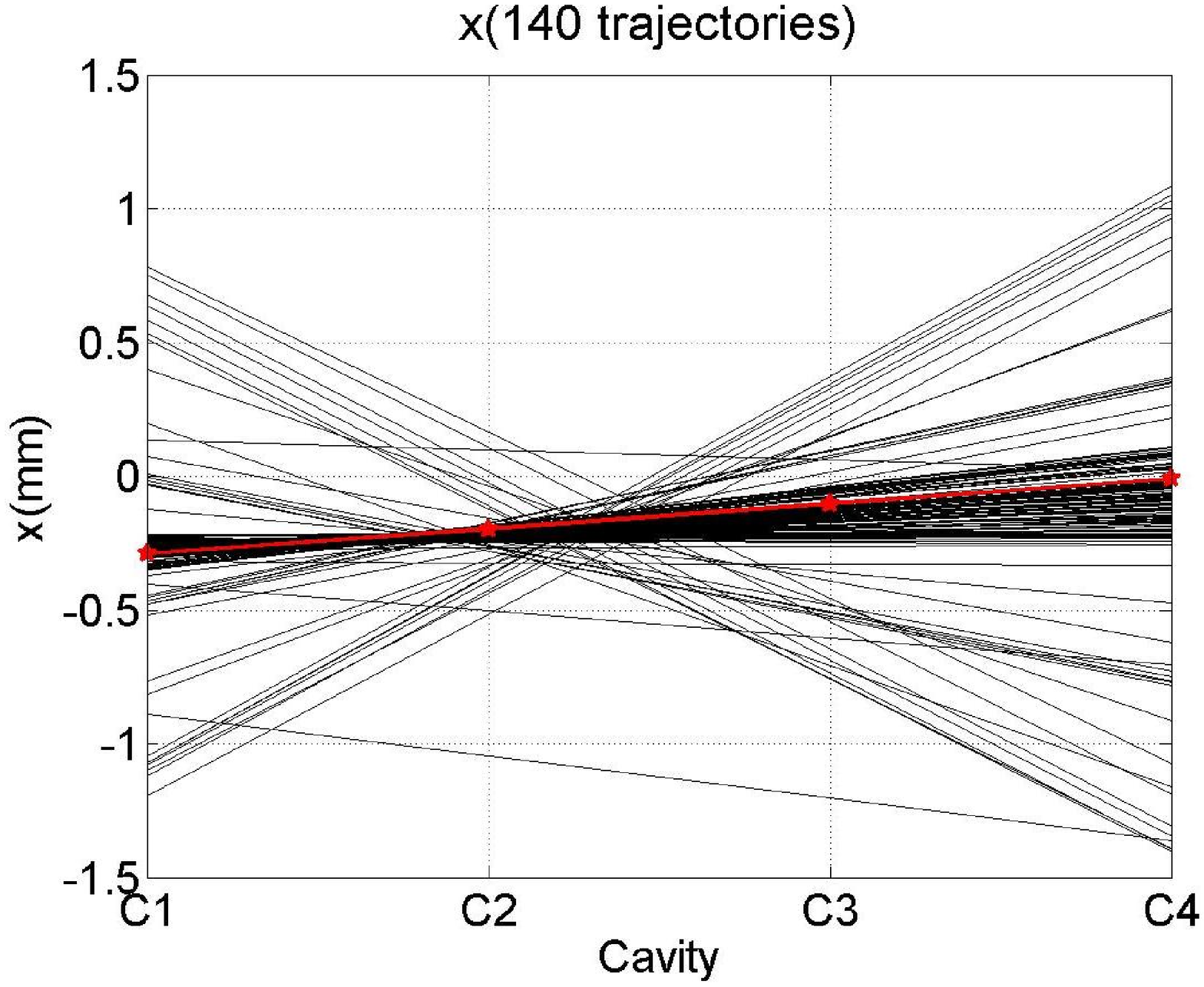}
\label{all-track-x}
}
\subfigure[y]{
\includegraphics[width=0.226\textwidth]{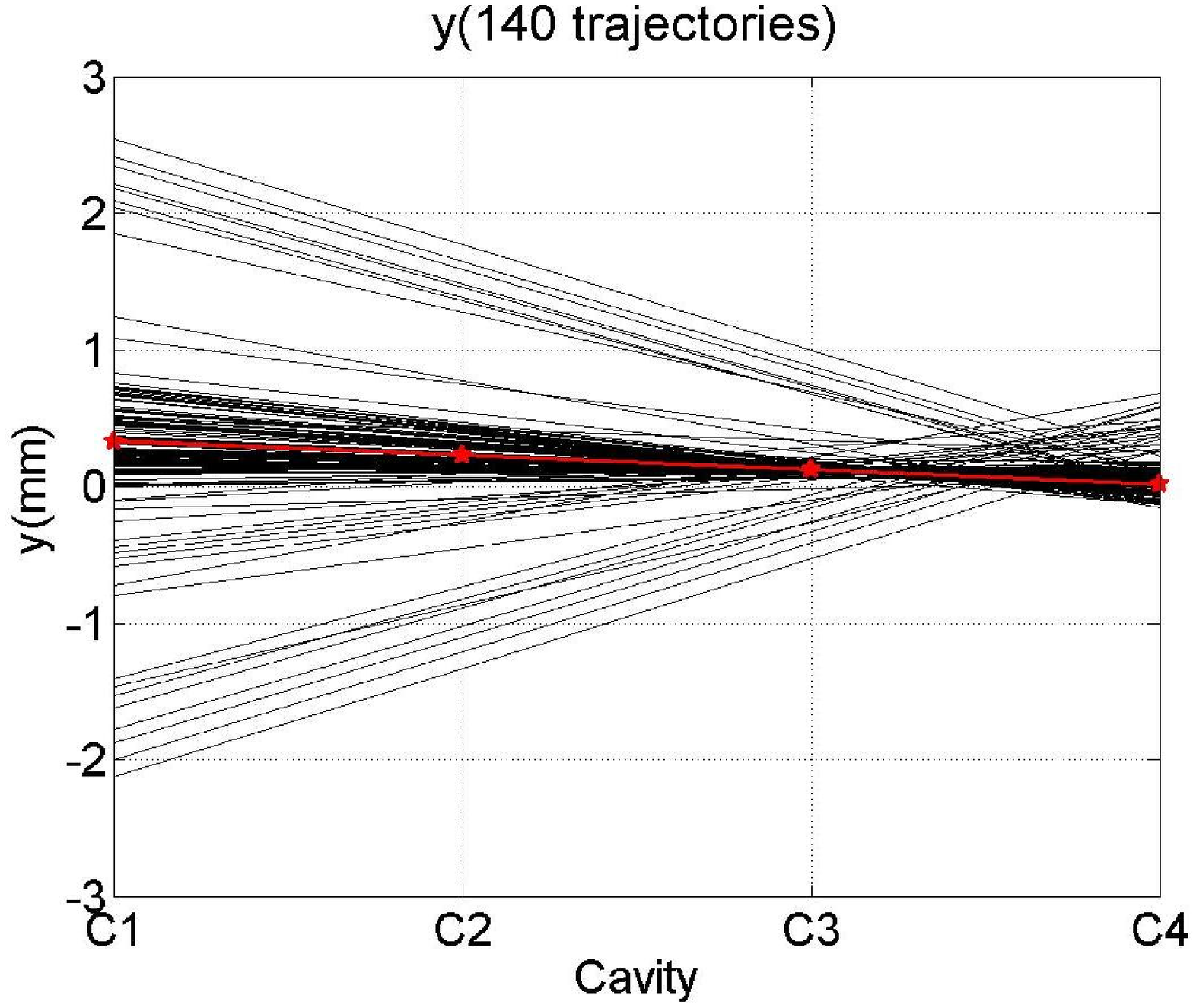}
\label{all-track-y}
}
\caption{Beam trajectories during movement. The red line marked with pentagram is the optimal trajectory based on coupler C2H2.}
\label{all-track}
\end{figure}

All beam trajectories across the four-cavity module during the scan are shown in Fig.~\ref{all-track}. Limited by the beam movement, not all possible trajectories can be covered. According to the position with minimum HOM power, the optimal beam trajectory based on coupler C2H2 is plotted as the red line. The red pentagrams located on C2 correspond to the position marked in Fig.~\ref{scan-xy} by red pentagram. 

\begin{figure}[h]
\centering
\subfigure[Optimal trajectory(x)]{
\includegraphics[width=0.226\textwidth]{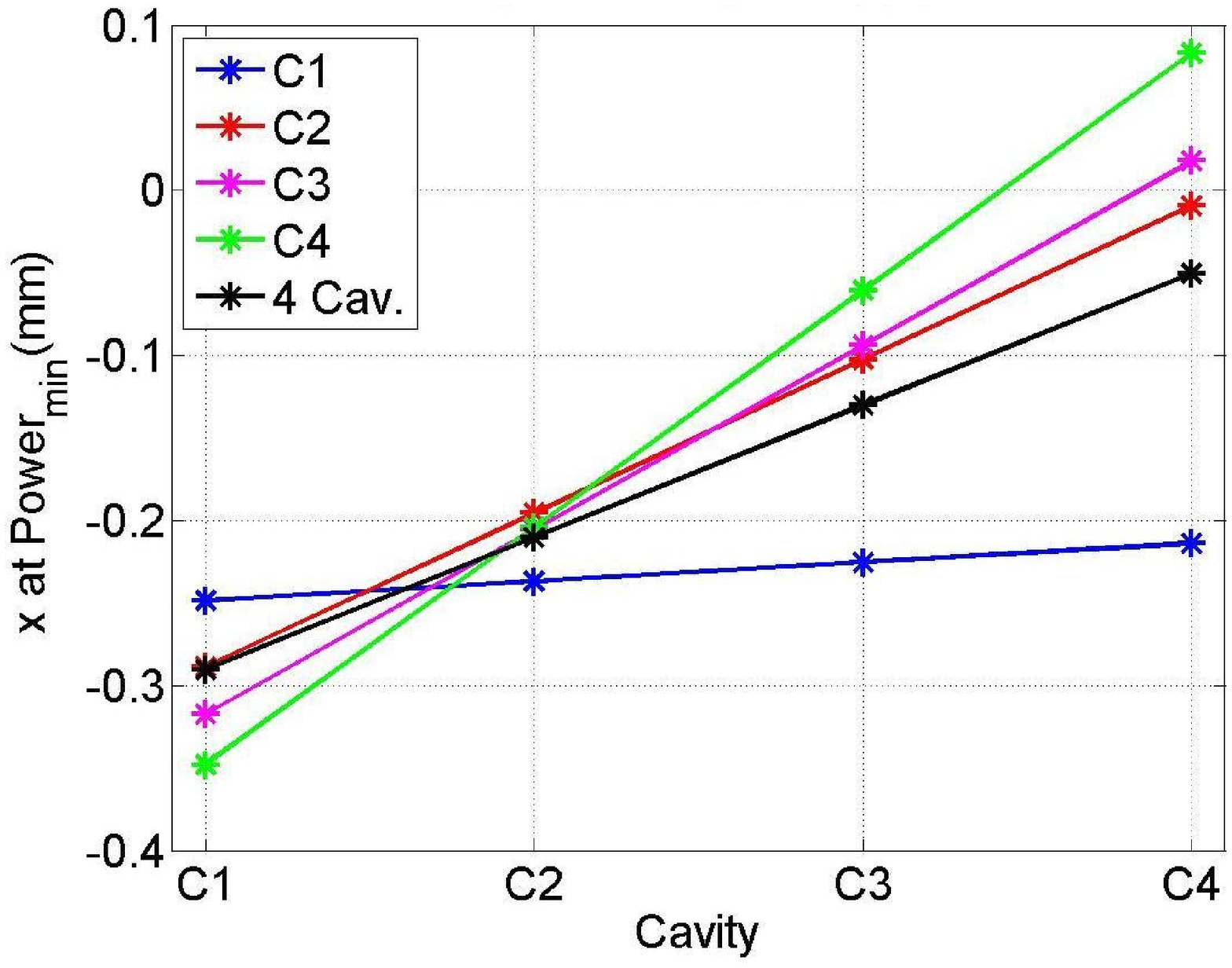}
\label{opt-track-x}
}
\subfigure[Optimal trajectory(y)]{
\includegraphics[width=0.226\textwidth]{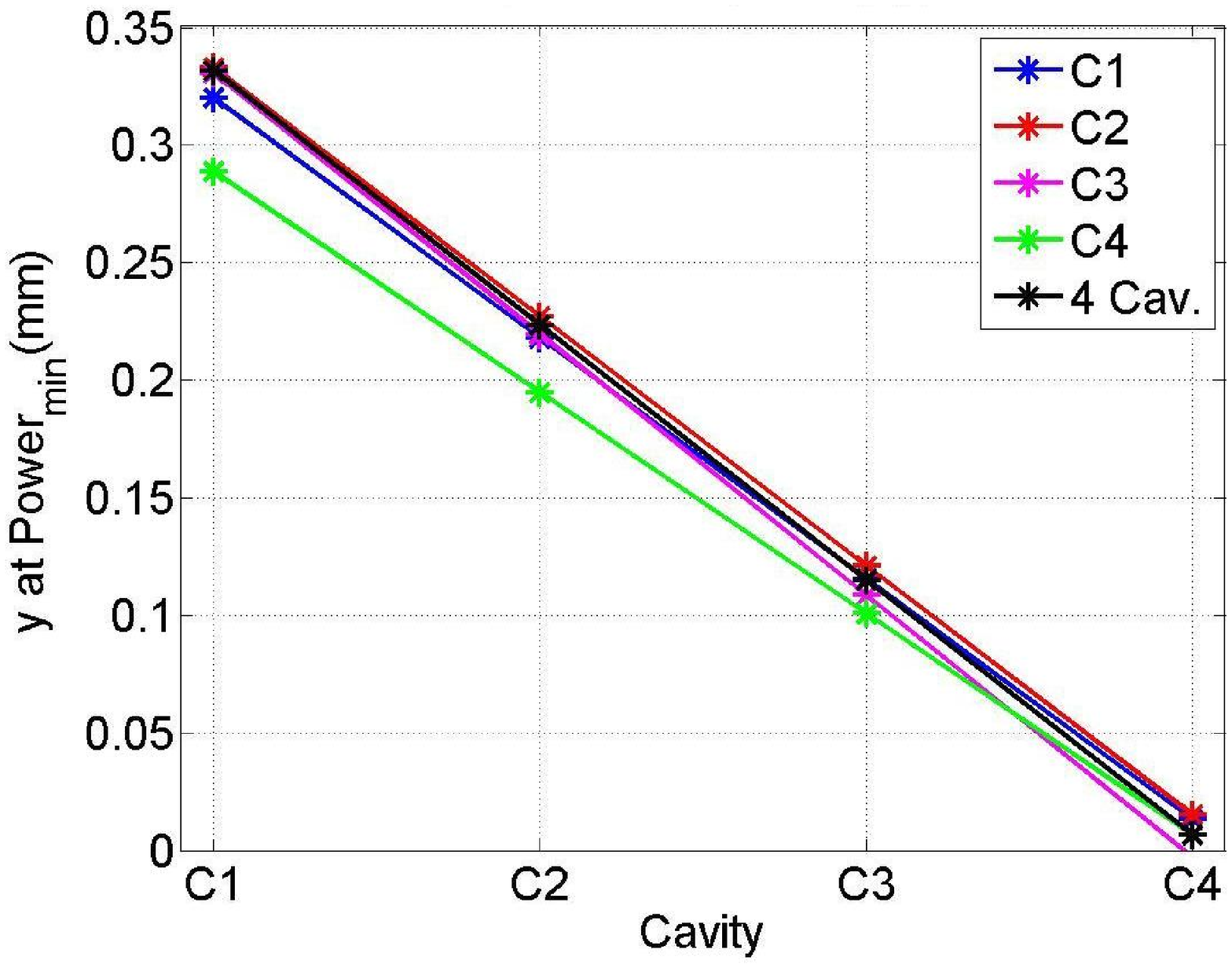}
\label{opt-track-y}
}
\caption{Optimal trajectories.}
\label{opt-track}
\end{figure}

As the HOM signals were also measured from the other three couplers for each beam movement, one can also f{}ind the position which has minimum power at each of them. The optimal trajectory for each coupler is shown in Fig.~\ref{opt-track}. Another trajectory is also obtained by minimizing the total power of all four couplers, which is the black line in Fig.~\ref{opt-track}. The optimal trajectories dif{}fer, and might be attributed to several factors. F{}irst, we could not cover the entire 4D space during the scan, therefore, the real optimal trajectory might have not been found in this study. Second, the BPMs used for position interpolations have f\mbox{}inite resolutions. Third, the alignment was based on a combination of several modes, and they may behave dif{}ferently at each coupler. Fourth, the HOM energy picked up by each coupler varies according to the detailed features of the individual couplers. Besides that, in spite of the related modes are propagating, dif{}ferent local f{}ield distributions are expected at each coupler. Therefore the HOM power is related to the overall pattern of the trajectory rather than the individual of{}fset in each cavity.
\begin{figure}[h]
\centering
\subfigure[norm(140 trajectories)]{
\includegraphics[width=0.226\textwidth]{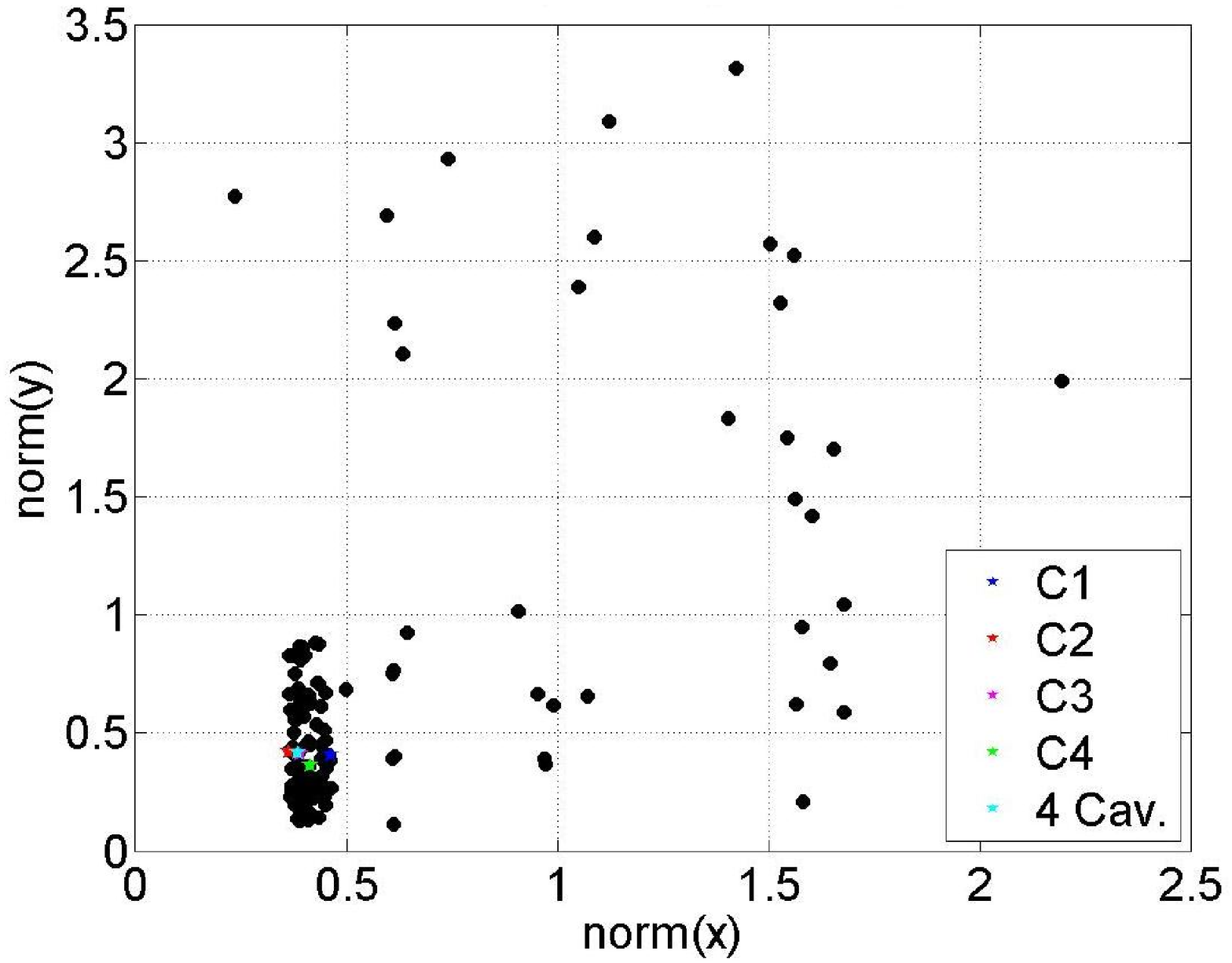}
\label{norm-all-track}
}
\subfigure[norm(optimal trajectories)]{
\includegraphics[width=0.226\textwidth]{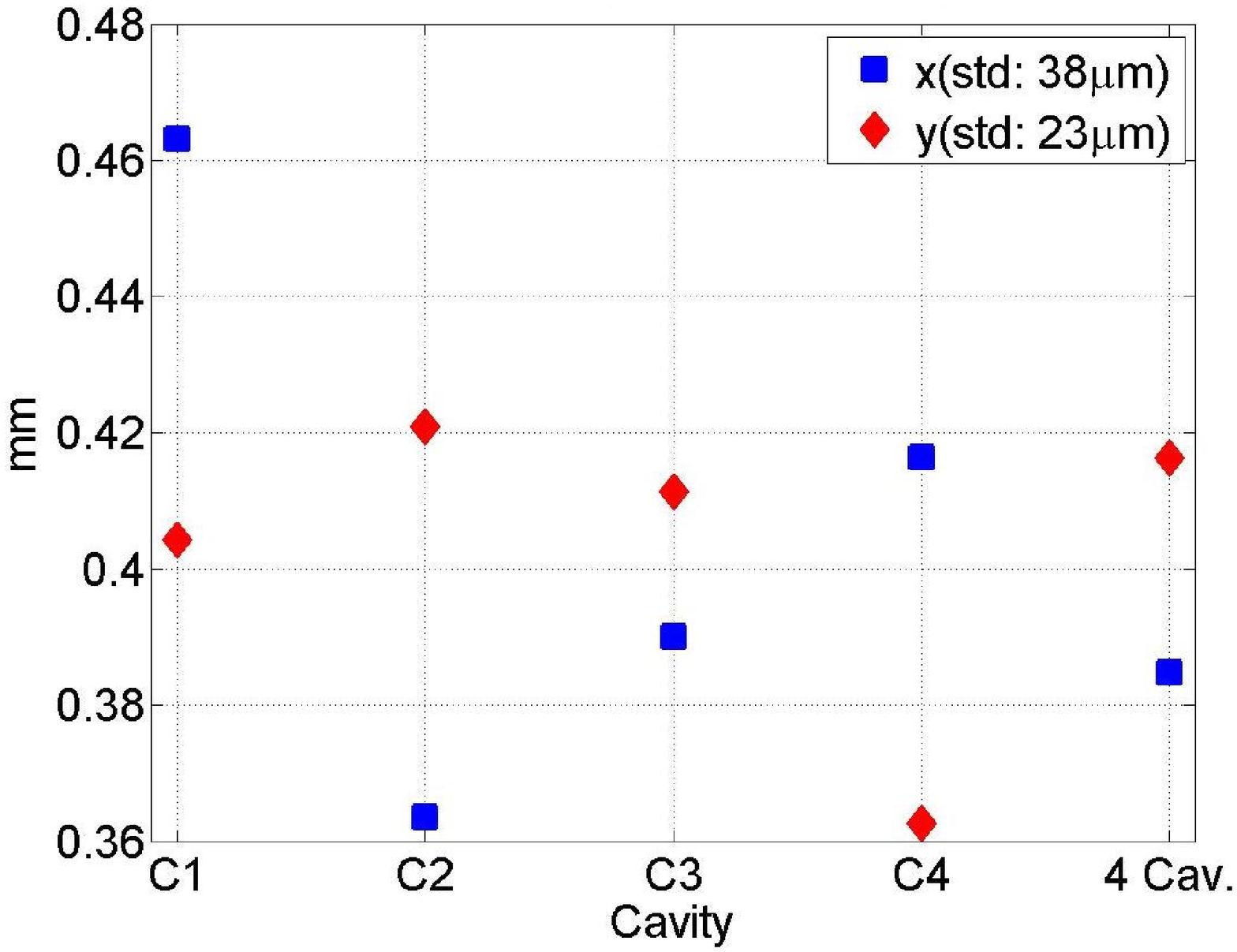}
\label{norm-opt-track}
}
\caption{Norm of all beam trajectories (a) and the optimal beam trajectories (b).}
\label{norm-track}
\end{figure}

In order to compare trajectories, a vector $\mathbf{A_i}$ is composed for each trajectory as
\begin{equation}
\mathbf{A_i} = (a^{(1)}_i, a^{(2)}_i, a^{(3)}_i, a^{(4)}_i)^T,
\label{eq:track-vector}
\end{equation}
where $a^{(1)}_i$, $a^{(2)}_i$, $a^{(3)}_i$ and $a^{(4)}_i$ denote the transverse beam position $x$ and $y$ of the $i^{th}$ trajectory in C1, C2, C3 and C4 respectively, $T$ denotes matrix transpose. Then the norm of each trajectory is calculated as
\begin{equation}
||\mathbf{A_i}|| = \sqrt{(a^{(1)}_i)^2+(a^{(2)}_i)^2+(a^{(3)}_i)^2+(a^{(4)}_i)^2}.
\label{eq:norm-vector}
\end{equation}
Since the modes excited by the beam at an angle are much weaker than those excited by the beam at an of{}fset \cite{racc1}, the angle of each trajectory is neglected in comparison. Fig.~\ref{norm-all-track} shows the norm of all trajectories and the f{}ive optimal trajectories. In terms of beam position $x$ and $y$, the norm of the f{}ive optimal trajectories are shown in Fig.~\ref{norm-opt-track}. The variations of optimal trajectories in both $x$ and $y$ directions are comparable to the RMS resolution of the two BPMs (30~$\mu$m).  

\section{Conclusions}\label{sec:conclude}
Beam alignment has been performed by minimizing the integrated power of dipole modes over a 30~MHz frequency range in the second dipole passband. These modes have been shown in simulations and measurements to have strong coupling to the beam and propagate across the four-cavity module. The beam trajectory corresponding to the minimum HOM power has been found based on each of the four downstream HOM couplers. The dif{}ferences among their norms are comparable to the BPM resolution. The overall alignment achieved based on one HOM coupler is of the order of 200~$\mu$m in this study.

Dedicated electronics for HOM-based beam diagnostics are currently under design. Once the HOM signal been calibrated, the position information within each cavity can be obtained directly. Therefore the nonlinearity induced by jitters and couplings of the BPMs might be reduced and consequently improve the beam alignment. We also plan to continue this study with the dedicated electronics.



\begin{thebibliography}{9} 

\bibitem{rwake} K.L.F.~Bane, SLAC-PUB-4169, 1986.

\bibitem{racc1-2} N.~Baboi, \emph{et al.}, LINAC2004, MOP36.

\bibitem{rflash} S.~Schreiber, \emph{et al.}, FEL2011,TUPB04.

\bibitem{racc39} E.~Vogel, \emph{et al.}, IPAC2010, THPD003.

\bibitem{rcst} CST Microwave Studio\textregistered, Ver.~2011, CST AG, Germany.

\bibitem{rdipac} P.~Zhang, \emph{et al.}, DIPAC2011, MOPD17.

\bibitem{racc1} S.~Molloy, \emph{et al.}, \emph{Phys. Rev. ST-AB} \textbf{9}, 112802 (2006).


\end{thebibliography}
\end{document}